\documentclass[pre,aps,onecolumn,showpacs]{revtex4}
\usepackage{amsmath,amssymb,graphicx,cases}
\usepackage{epsfig}
\usepackage{textcomp}

\begin{document}
\title{The number statistics and optimal history of non-equilibrium steady states of mortal diffusing particles}
\author{Baruch Meerson}
\affiliation{Racah Institute of Physics, Hebrew University of
Jerusalem, Jerusalem 91904, Israel}


\begin{abstract}
Suppose that a point-like steady source at $x=0$ injects particles into a half-infinite line. The particles diffuse and die. At long times a non-equilibrium steady state sets in, and we assume that it involves many particles. If
the particles are non-interacting, their total number $N$ in the steady state is Poisson-distributed with mean $\bar{N}$ predicted from a deterministic reaction-diffusion equation. Here we determine the most likely density history of this driven system conditional on observing a given $N$.
We also consider two prototypical examples of \emph{interacting} diffusing particles: (i) a family of mortal diffusive lattice gases with constant diffusivity (as illustrated by the simple symmetric exclusion process with mortal particles), and (ii) random walkers that can annihilate in pairs. In both examples we calculate the variances of the (non-Poissonian) stationary distributions of $N$.

\end{abstract}
\maketitle

\section{Introduction}

Fluctuations of non-equilibrium steady states of driven diffusive lattice gases have attracted a lot of attention in the last two decades \cite{Spohn,KL99,SZ95,S00,D07,BE07,JL2014}. Although many of these studies assumed purely diffusive, particle conserving dynamics, lattice gas models with dissipation have also been investigated \cite{JonaLasinio93,Basile,BodineauLag,Hurtado}.  The absence of detailed balance makes dissipative
systems more difficult to handle.  There is, however, a steady interest in dissipative models, mostly because of their relevance to experiment in such diverse areas as fluid turbulence \cite{Frisch}, granular gases \cite{Poeschel1,Poeschel2} and many other non-equilibrium settings in physics, chemistry, biology and engineering.

The simplest way to characterize a non-equilibrium system is to study its steady state. To maintain a dissipative systems in a steady state one must constantly drive it by injecting energy or mass.  The total energy or mass content of the driven system
fluctuates around the mean, and these fluctuations bear the stamp of the non-equilibrium nature of the system. In this work we consider three different driven models of diffusive particles where particles can die: either individually or by annihilating in pairs. In each of these models the system is driven by injecting particles into a half-infinite straight line from a single point-like steady source. The particles diffuse and die so that,  at long times, a non-equilibrium steady state sets in. If the particles do not interact, their total number $N$ is Poisson-distributed with mean $\bar{N}$ predicted from a deterministic reaction-diffusion equation for this driven system. Our main interest in this case will be to find the optimal (that is,  most probable) density history of the system conditional on observing a given $N$. We also consider two types of \emph{interacting} particle models: (i) diffusive lattice gases of mortal particles with constant diffusivity but non-trivial fluctuations (as illustrated, for example, by the simple symmetric exclusion process \cite{Spohn} with mortality), and random walkers on a lattice that can annihilate in pairs. In both these cases the statistics of $N$ is expected to be non-Poissonian, and we calculate the \emph{variances} describing typical, Gaussian fluctuations of the particle number around the mean. We obtain these results by employing (a dissipative extension of) the Macroscopic Fluctuation Theory (MFT): a coarse-grained low-noise large-deviation theory that employs, as a large parameter, the typical number of particles in the region of interest,  see Ref. \cite{MFTreview} for a recent review. The applicability of the MFT in the driven systems, considered in this work, demands $\bar{N}\gg 1$, and we will work in the parameter regions where this condition is satisfied.

Here is a plan of the remainder of the paper. Section II starts with a brief exposition of the expected, or average behavior of a driven system of non-interacting random walkers or Brownian particles
that die individually. Then subsection IIB presents the MFT formulation of the problem of particle number statistics for a more general family of diffusive lattice gases of particles that die individually. For the non-interacting particles, we are able to solve,
in the same Subsection IIB, the MFT equations exactly. The solution, via the Hopf-Cole transformation, reproduces (the large-$N$ asymptotic of) the expected Poisson distribution of $N$. In addition, it gives the previously unavailable optimal density history of the driven system, conditional on observing a given $N$.  Section III deals with two examples of \emph{interacting} particles.  The first of them deals with a family of mortal interacting diffusive lattice gases with constant diffusivity but non-trivial fluctuations, as illustrated by the simple symmetric exclusion process with mortal particles. The second example involves random walkers that only interact via pair-wise annihilation. In these examples a full solution of the MFT problem  is presently unavailable, and we only calculate the variances of the respective stationary distributions of $N$. Our main results and their possible extensions are briefly discussed in Section IV.

\section{Mortal random walkers}

\subsection{Expected density behavior}

Consider a half-infinite one-dimensional lattice with lattice constant $a$ and
suppose that a source of particles at the origin, $x=0$, sets a constant particle number density $n_0$ there. The particles perform random walk at $x>0$ with diffusivity
$D$ and die individually with rate $\mu$. When $D\gg \mu a^2$, there is no difference between the discrete random walk and continuous diffusion, and the average particle density $\rho(x,t)$ is governed by the reaction-diffusion equation \cite{Spohn,Lebowitz,Mikhailov}
\begin{equation}\label{MFeq}
\partial_t \rho = -\mu \rho + D\partial_x^2 \rho,\;\;\;\;\;0<x<\infty.
\end{equation}
At long times the average density profile approaches a steady state, independent of the initial condition:
\begin{equation}\label{MFsteadysol}
\bar{\rho}(x) = n_0 e^{-\sqrt{\frac{\mu}{D}} x}.
\end{equation}
Correspondingly, the average steady-state number of particles in this driven system is
\begin{equation}\label{Nbarsteady}
\bar{N} =\int_0^{\infty} dx\,\bar{\rho}(x) = \sqrt{\frac{D}{\mu}} \,n_0,
\end{equation}
and we assume that this number is much larger than unity. The actual number of particles $N$ in the steady state fluctuates around $\bar{N}$ because of the shot noises of the diffusion and mortality. We are interested in the steady-state probability distribution of $N$. For non-interacting random walkers that die individually this probability distribution can be found exactly, by solving the steady-state master equation for the multi-variate probability distribution of observing $n_1$ particles on site 1, $n_2$ particles on site 2, etc. The solution has the form of the Poisson product measure with space-dependent parameters [for $D\gg \mu a^2$, this measure corresponds to the average density profile (\ref{MFsteadysol})].  This leads to a Poisson distribution of $N$ in the steady state. We will proceed, however, as if we were unaware of these exact results, and employ instead the MFT: a coarse-grained low-noise theory mostly based on the strong inequality $\bar{N}\gg 1$.  The purpose is two-fold. First,  even for the non-interacting particles, the MFT will give the previously unknown optimal density history of the driven system, conditional on observed $N$. The optimal density history is well defined only in the limit of  $\bar{N}\gg 1$, where it is \emph{much} more likely than other histories leading to the same $N$. Second, our main interest is in \emph{interacting} particle models, where exact microscopic results are usually unavailable. In the next subsection we briefly discuss  the basics of the  MFT, and formulate the MFT problem for the particle number statistics in a broader context of a family of driven lattice gases of mortal particles.

\subsection{Macroscopic Fluctuation Theory (MFT) of the Particle Number Statistics}

\subsubsection{Governing Equations}

When (i) the length scale of interest is much larger than the lattice constant $a$, (ii) the time scale of interest is much larger than the inverse rates of the microscopic processes of diffusion and death, and (iii) the typical number of involved particles is much larger than unity, the statistics of large deviations is captured by the MFT \cite{JonaLasinio93,Basile,EK,BodineauLag,MS2011,MFTreview}.  The starting point of the derivation of the MFT for diffusing and reacting particles is the exact master equation for the multi-variate probability distribution of observing a certain number of particles on each cite. Here one can either work directly in the physical space, or employ the multi-site probability generating function (that, in the spatially-continuous limit, becomes a probability generating functional). Going over to a path-integral formulation, one then makes a low-noise approximation by evaluating the path integral by the Laplace method that employs the number of particles in the relevant region of space as a large parameter. This procedure yields saddle-point equations (partial differential equations) that can be written in a Hamiltonian form: for the density field $q(x,t)$ and a conjugate field $p(x,t)$ that plays a role of the ``momentum density". At a qualitative level, the conjugate field $p(x,t)$ describes the magnitude of fluctuations.

If the calculations are performed in the physical space, the saddle point equations (presented here in a form, suitable
for a class of diffusive lattice gases of particles that die individually), take the form \cite{JonaLasinio93,Basile,BodineauLag,MS2011,MFTreview}
\begin{eqnarray}
  \partial_t q &=& -\mu q e^{-p}+\partial_x\left[D(q) \partial_x q-\sigma(q) \partial_x p\right], \label{d1gen} \\
  \partial_t p &=&  -\mu (e^{-p}-1)- D(q) \partial_x^2 p-\frac{1}{2} \,\sigma^{\prime}(q) (\partial_x p)^2, \label{d2gen}
\end{eqnarray}
where $D(q)$ is the gas diffusivity,  $\sigma(q)$ is (twice) the mobility  \cite{Spohn}, and the prime denotes the derivative with respect to the argument. The terms proportional to $\mu$
describe the on-site particle death and its fluctuations. The rest of terms describe diffusive transport and its fluctuations.
Equations (\ref{d1gen}) and (\ref{d2gen}) are indeed Hamiltonian, as they can be written in terms of
variational derivatives:
\begin{equation}
\partial_t q = \delta H/\delta p\,, \quad
\partial_t p = -\delta H/\delta q\,,
\end{equation}
where
\begin{equation}
\label{Hamiltonian}
H\{q(x,t),p(x,t)\}= \int_0^{\infty} dx\,\mathcal{H}
\end{equation}
is the Hamiltonian, and
\begin{equation}
\label{Ham}
\mathcal{H}(q,p) = \mu q\left(e^{-p}-1\right) -D(q) \partial_x q \partial_x p
+\frac{1}{2}\sigma(q)\!\left(\partial_x p\right)^2
\end{equation}
is the Hamiltonian density. Going back to the non-interacting mortal random walkers, we put $D(\rho)=D=\text{const}$ and $\sigma(\rho)=2D\rho$, see e.g. Ref. \cite{Spohn}. Then  Eqs.~(\ref{d1gen}) and (\ref{d2gen}) become
\begin{eqnarray}
  \partial_t q &=& -\mu q e^{-p} + D \partial_x\left(\partial_x q-2 q \partial_x p\right), \label{d1} \\
  \partial_t p &=& -\mu (e^{-p}-1)- D \partial_x^2 p-D (\partial_x p)^2, \label{d2}
\end{eqnarray}
where $0<x<\infty$. The boundary conditions at the particle source are $q(x=0,t)=n_0$ and $p(x=0,t)=0$ \cite{MS2011}. The latter condition is quite intuitive: as we demand a fixed (deterministic) value of the density at $x=0$, $p$ must vanish there.  Far away from the source there are no particles. This brings the boundary condition $q(x=\infty, t)=0$.

Being interested in steady state fluctuations, we can assume that, at $t=-\infty$, the system is at its deterministic steady state \cite{EK,Kurchan}: $q(x,t=-\infty)= \bar{\rho}(x)$, see Eq.~(\ref{MFsteadysol}). We condition the process on observing $N$ particles at some finite moment of time that, without loss of generality, we can set to zero. This imposes an \emph{integral} constraint on the solution at $t=0$:
\begin{equation}\label{number}
\int_0^{\infty} dx\, q(x,t=0) = N.
\end{equation}
Analogous integral constraints appear in the MFT formulations of the problem of statistics of integrated current in an infinite setting \cite{DG2009b} and statistics of particle absorption by an absorber at $x=0$ \cite{MR}. To account for the integral constraint, we should introduce a Lagrange multiplier $\lambda$ and minimize the extended action that incorporates the integral constraint.  Similarly to Refs. \cite{DG2009b} and \cite{MR}, the action minimization does not change the ``bulk" equations \eqref{d1gen} and \eqref{d2gen} [or \eqref{d1} and \eqref{d2}], but yields an additional boundary condition for $p(x,t)$ at $t=0$:
\begin{equation}\label{pT}
    p(x,t=0)=\lambda \, \theta(x),
\end{equation}
where $\theta(x)$ is the Heaviside step function, and $\lambda$ is ultimately set by Eq.~(\ref{number}) \cite{DG2009b,MR}.

Note that $p(x,t)=0$ is an invariant manifold of Eqs.~(\ref{d1gen}) and (\ref{d2gen}). The dynamics on this manifold
is described by the deterministic Eq.~(\ref{MFeq}). This is the \emph{relaxation} path of the system; it solves the problem in
the particular case $N=\bar{N}$. For $N\neq \bar{N}$, the solution of the MFT equations describes the optimal \emph{activation} path: the most likely density history of the driven system conditional on observing $N$ particles. Here $p(x,t)\neq 0$. Once $q(x,t)$ and $p(x,t)$ are found, we can evaluate the action $S$ that yields ${\mathcal P} (N)$  up to a pre-exponential factor:
\begin{eqnarray}
\label{actionmain}
  &-&\ln {\mathcal P}(N) \simeq S = \int_{-\infty}^0 dt\, \int_0^{\infty} dx\,\left(p\partial_t q-\mathcal{H}\right) \nonumber\\
  &=& \int_{-\infty}^0  dt\, \int_0^{\infty} dx\,\left[Dq(\partial_x p)^2+\mu q \left(1-e^{-p}-pe^{-p}\right)\right].
\end{eqnarray}
The first term of the integrand comes from the shot noise of diffusion, the second term comes from  the shot noise of mortality. Rescaling time $\mu t \to t$, the coordinate $\sqrt{\mu/D}\, x \to x$ and the density $q/n_0 \to q$, one can see
that $-\ln {\mathcal P}(N)$ obeys a simple scaling relation
\begin{equation}\label{scalingRW}
-\ln {\mathcal P}(N) = \bar{N} \,f\left(\frac{N}{\bar{N}}\right),
\end{equation}
where $f(z)$ is the large deviation function of the number of particles. Note that
$n_0$ only enters this scaling relation through $\bar{N}$.

\subsubsection{Particle number statistics and optimal path}

We note that Eq.~(\ref{d2}) is decoupled from Eq.~(\ref{d1}). This decoupling only occurs for non-interacting particles, and it greatly simplifies the problem. Let us perform the Hopf-Cole transformation by introducing $Q=q e^{-p}$ and $P=e^p-1$ \cite{EK}. The generating functional of this canonical transformation can be chosen to be
\begin{equation}\label{F}
F\{q(x,t),Q(x,t)\}=\int_0^\infty dx\, \left(q \ln \frac{q}{Q}-q+Q\right).
\end{equation}
In the new variables $Q$ and $P$ the Hamiltonian is $\tilde{H}\{Q(x,t),P(x,t)\}=\int dx\,\tilde{{\mathcal H}}$, where
$$
\tilde{{\mathcal H}} = -\mu QP -D\, \partial_x Q \,\partial_x P.
$$
As a result, the MFT equations become linear and fully decoupled:
\begin{eqnarray}
  \partial_t Q &=& -\mu Q + D\partial_x^2 Q, \label{Qt}\\
  \partial_t P &=& \mu P - D\partial_x^2 P. \label{Pt}
\end{eqnarray}
Note that these equations for $Q$ and $P$ arise immediately, when one employs the Laplace method for the evaluation of the path integral in the formalism of probability generating functional \cite{EK}.

In the new variables $Q$ and $P$, the boundary and initial conditions are:
\begin{equation}\label{forQ}
Q(0,t)=n_0 \;\;\; \mbox{and}\;\;\;Q(x,-\infty)=\bar{\rho}(x)
\end{equation}
for $Q$, and
\begin{equation}\label{forP}
P(0,t)=0\;\;\; \mbox{and}\;\;\;P(x,0)=\left(e^{{\lambda}}-1\right)\,\theta(x)
\end{equation}
for $P$. As a result, $Q(x,t)$ is invariant in time,
\begin{equation}\label{Qsteady}
    Q(x,-\infty<t\leq 0) = \bar{\rho}(x),
\end{equation}
while
\begin{equation}\label{Psolsteady}
   P(x,-\infty<t\leq 0)=\left(e^{\lambda}-1\right)\,e^{\mu t}\,\text{erf}\,\left(\frac{x}{\sqrt{-4D t}}\right).
\end{equation}
Now we can determine the optimal path in the original variable $q$:
\begin{equation}
\label{optimalsteady}
  q(x,t) = Q(x) [1+P(x,t)] = n_0 e^{-\sqrt{\frac{\mu}{D}} x} \left[1+\left(e^{\lambda}-1\right)\,e^{\mu t}\,\text{erf}\,\left(\frac{x}{\sqrt{-4D t}}\right)\right].
\end{equation}
Using Eq.~(\ref{number}), we find $\lambda=\ln (N/\bar{N})$, so
\begin{equation}
\label{optimalsteadyfinal}
  q(x,t) = n_0 e^{-\sqrt{\frac{\mu}{D}} x} \left[1+\left(\frac{N}{\bar{N}}-1\right)\,e^{\mu t}\,\text{erf}\,\left(\frac{x}{\sqrt{-4D t}}\right)\right].
\end{equation}
It is easier to  calculate the action in the new variables $Q$ and $P$ where, as one can show by a direct calculation  \cite{MR},  the action is equal to the increment of generating functional $F$ from Eq.~\eqref{F}:
\begin{equation}\label{increment}
S=F\{q(x,0), Q(x,0)\}-F\{q(x,-\infty), Q(x,-\infty)\}.
\end{equation}
After some algebra, this gives
\begin{eqnarray}
- \ln {\mathcal P}(N) \simeq S &=& \int_{0}^{\infty} dx\,\left[q(x,0) \ln\frac{q(x,0)}{Q(x,0)}-q(x,0)+Q(x,0)\right] \nonumber\\
  &=& N \ln\frac{N}{\bar{N}}-N+\bar{N},
  \label{Poisson}
\end{eqnarray}
where $\bar{N}$ is given by Eq.~(\ref{Nbarsteady}). That is, the large deviation function $f(z)$ from Eq.~(\ref{scalingRW}) is equal to $f(z)=z \ln z-z+1$.  The distribution (\ref{Poisson}) coincides with the $N\gg 1$, $\bar{N}\gg 1$ asymptotic of the Poisson distribution with mean $\bar{N}$, as to be expected. In particular, the variance of this distribution coincides with the mean:
\begin{equation}\label{varRW}
    V_{\text{RW}}=\bar{N}=\sqrt{\frac{D}{\mu}}\,n_0.
\end{equation}

\begin{figure}
\includegraphics[width=0.4\textwidth,clip=]{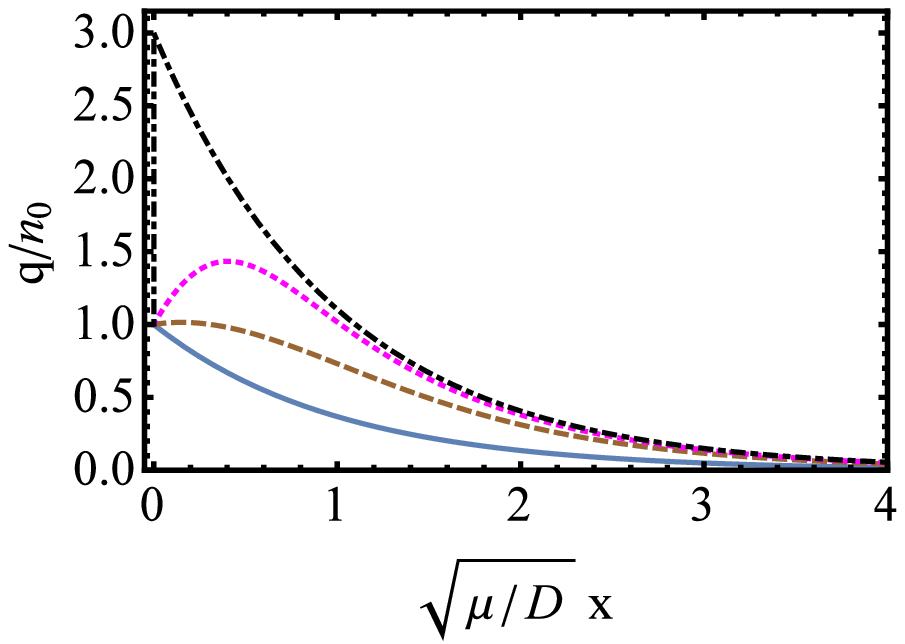}
\includegraphics[width=0.4\textwidth,clip=]{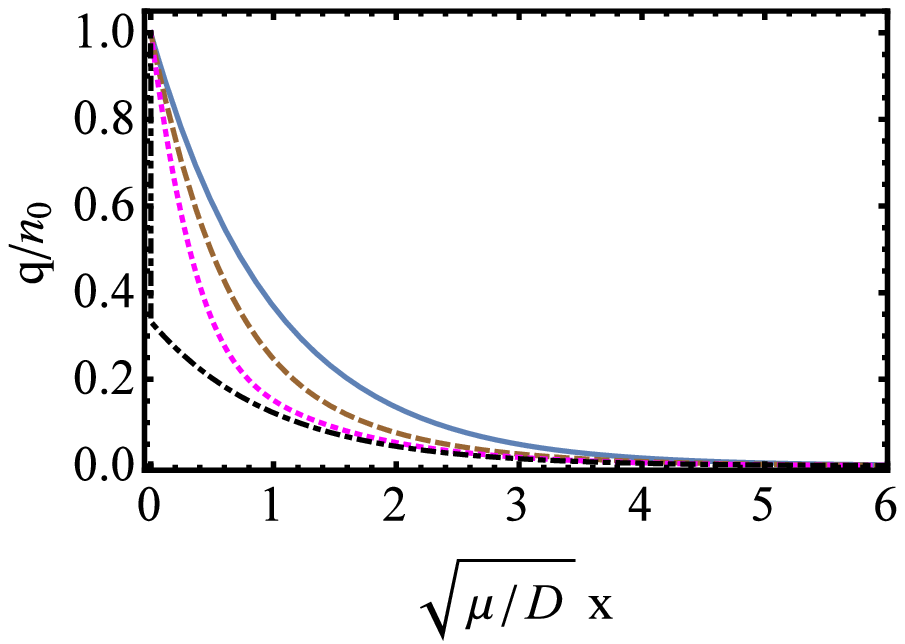}
\caption{(Color online) The optimal density history for $N/\bar{N}=3$  (the left panel) and $N/\bar{N}=1/3$ (the right panel). The times, rescaled by the decay rate $\mu$, for both panels,  are $-\infty$ (solid line), $-0.4$ (dashed line), $-0.1$ (dotted line) and $0$ (dash-dotted line). At $t=0$ a density jump develops at the origin, so that the required number of particles $N$ effectively comes from a deterministic density profile with a different density at the source.}
\label{1}
\end{figure}

Now let us return to the optimal path (\ref{optimalsteadyfinal}) that has been previously unknown. Although the statistics of $N$ is time-independent,  the optimal path does depend on time. Furthermore, the activation path does not coincide with the time-reversed relaxation path, obtained by solving the deterministic reaction-diffusion equation
(\ref{MFeq}) back in time.  This is a clear signature of non-equilibrium. Notice also that, in order to ensure an unusually large or small number of particles at $t=0$, the fluctuations create a boundary layer in the density profile at the particle source. This boundary layer  becomes a density jump at $t=0$,
\begin{equation}\label{t=0}
q(x>0,t=0)= \frac{N}{\bar{N}}\, \bar{\rho}(x),
\end{equation}
so that the \emph{effective} boundary condition is $q(x\to 0,t=0)=n_0 N/\bar{N}$, whereas the bulk of the gas particles behaves deterministically. These features can be seen on Figure \ref{1} which shows the optimal density histories described by Eq.~(\ref{optimalsteadyfinal}).
The left and right panels correspond to $N=3\bar{N}$ and $N=\bar{N}/3$, respectively. These results are both unexpected and instructive.

\section{Interacting lattice gases}
\label{var1}

Now let us consider interacting lattice gases of mortal particles. They may have different $\sigma(q)$, but for simplicity we will
continue to assume a constant diffusivity $D=\text{const}$. For such gases the deterministic equations~(\ref{MFeq})-(\ref{Nbarsteady}) continue to hold, while the MFT equations read
\begin{eqnarray}
  \partial_t q &=& -\mu q e^{-p} + \partial_x \left[D\partial_x q-\sigma(q) \partial_x p\right], \label{d3gen} \\
  \partial_t p &=& -\mu (e^{-p}-1)- D \partial_x^2 p-\frac{1}{2}\sigma^{\prime}(q) (\partial_x p)^2,\label{d4gen}
\end{eqnarray}
with the same boundary conditions as before.  Once $q(x,t)$ and $p(x,t)$ are known, the probability distribution  ${\mathcal P} (N)$ can be evaluated from
\begin{eqnarray}
\label{actionsigma}
  &-&\ln {\mathcal P}(N) \simeq S = \int_{-\infty}^0  dt\, \int_0^{\infty} dx\,\left[\frac{1}{2}\sigma(q) (\partial_x p)^2+\mu q \left(1-e^{-p}-pe^{-p}\right)\right].
\end{eqnarray}
Rescaling time $\mu t \to t$ and the coordinate $\sqrt{\mu/D} \,x \to x$, we obtain a scaling relation
\begin{equation}\label{scalingsigma}
-\ln {\mathcal P}(N) = \bar{N} \,f\left(\frac{N}{\bar{N}},n_0\right),
\end{equation}
where $n_0$ enters both through $\bar{N}$ and separately \cite{scalesigma}.

It does not seem possible to solve Eqs.~\eqref{d3gen} and \eqref{d4gen} and determine the large deviation function $f(z,n_0)$ analytically for a general $\sigma(q)$. Typical, Gaussian fluctuations of the number of particles around $\bar{N}$ are given by a quadratic asymptotic of $f(z)$ at $z$ close to $1$. This asymptotic can be found relatively easily via a perturbation theory around the deterministic steady-state solution (\ref{MFsteadysol}). This theory employs the Lagrange multiplier $\lambda$ as a small parameter \cite{KM_var}. We set
\begin{subequations}
\begin{align}
\label{q_exp}
q &= \bar{\rho}(x)+ \lambda q_1 + \lambda^2 q_2+\ldots \\
\label{p_exp}
p &= \qquad \quad\;\;\; \lambda p_1 + \lambda^2 p_2+\ldots
\end{align}
\end{subequations}
and plug these expansions into Eqs.~\eqref{d3gen} and \eqref{d4gen}. The first-order equations are
\begin{subequations}
\begin{align}
\label{q1}
&(\partial_t +\mu -D\partial^2_{x}) \,q_1 =  \mu  \bar{\rho}(x) p_1-\partial_x[\sigma(\bar{\rho})\partial_x p_1],\\
\label{p1}
&(\partial_t -\mu +D \partial^2_{x}) \,p_1=0.
\end{align}
\end{subequations}
The equation for $p_1$ is independent of $\sigma(q)$, and it is decoupled from the equation for $q_1$. Therefore, we can solve it immediately, with the
boundary conditions $p_1(0,t)=0$ and $p_1(x,0)=\theta(x)$. The solution is
\begin{equation}\label{p1sol}
p_1(x,t\leq 0)=e^{\mu t} \,\text{erf}\,\left(\frac{x}{\sqrt{-4Dt}}\right).
\end{equation}
Now we could plug this expression in Eq.~(\ref{q1}) and solve for $q_1$ with the boundary conditions $q_1(0,t)=0$
and $q_1(x,-\infty)=0$. This is unnecessary, however, for the purpose of computing the variance of ${\mathcal P}(N)$,
because the latter is independent of $q_1$. Indeed, we have
\begin{eqnarray}
\label{actionvariance}
  &-&\ln {\mathcal P}(N) \simeq S = \frac{\lambda^2}{2}
 \int_{-\infty}^0  dt\, \int_0^{\infty} dx\,\left[\sigma(\bar{\rho}) (\partial_x p_1)^2
 +\mu \bar{\rho}p_1^2\right] +{\mathcal O}(\lambda^3).
\end{eqnarray}
Correspondingly, the variance is equal to
\begin{eqnarray}
\label{variance}
 V = \int_{-\infty}^0  dt\, \int_0^{\infty} dx\,\left[\sigma(\bar{\rho}) (\partial_x p_1)^2
 +\mu \bar{\rho}p_1^2\right],
\end{eqnarray}
with $\bar{\rho}(x)$ from Eq.~(\ref{MFsteadysol}) and $p_1(x,t)$ from Eq.~(\ref{p1sol}).  Equation~(\ref{variance}) does not demand a knowledge of $q_1(x,t)$ and holds for any mortal diffusive lattice gas with $D=\text{const}$.

One well-known example of a gas with constant diffusivity $D$ but non-trivial fluctuations is provided by the SSEP, where each
particle can randomly hope to a neighboring lattice site if
that site is vacant. If it is occupied by another particle, the
move is forbidden. For the SSEP one has $\sigma(\rho)=2D\rho (1-\rho)$ \cite{Spohn}. Here we set the lattice constant $a=1$,
so that the particle density at the source $0<n_0\leq 1$ is dimensionless.   Evaluating the double integral in Eq.~(\ref{variance})  in this case (see Appendix \ref{SSEP}),
we obtain
\begin{equation}\label{varianceSSEP}
V_{\text{SSEP}}= \sqrt{\frac{D}{\mu}} \,n_0 \left(1-\frac{2 n_0}{\pi}\right) = \bar{N} \left(1-\frac{2 n_0}{\pi}\right).
\end{equation}
As $V_{\text{SSEP}}\neq \bar{N}$, ${\mathcal P}(N)$ is non-Poissonian. As expected on the physical grounds, $V_{\text{SSEP}}$ is smaller than the variance of the total number of non-interacting random walkers with the same $n_0$, see Eq.~(\ref{varRW}), so the distribution is narrower than the Poisson distribution with the same mean. The two variances coincide in the limit of $n_0\to 0$, where exclusion effects in the SSEP are negligible. That ${\mathcal P}(N)$ is non-Poissonian is not surprising, but even its variance has been previously unknown.

\section{Annihilating Random Walkers}

An annihilating random walker (ARW) is immortal when it is alone, but two ARWs on the same lattice site can annihilate, $2A \to \emptyset$.  Let $\alpha$ be the annihilation rate constant. When the diffusion is sufficiently fast [see the criterion \eqref{ineq1} below]
the average particle density $\rho(x,t)$ is governed by the continuous reaction-diffusion equation \cite{MS2011}
\begin{equation}\label{MFeq1}
\partial_t \rho = -\alpha \rho^2 + D\partial_x^2 \rho,\;\;\;\;\;0<x<\infty.
\end{equation}
The particle source at $x=0$ fixes a particle density $n_0$ of the ARWs there. With this boundary condition, the steady-state average density profile is
\begin{equation}\label{MFsteadysol1}
\bar{\rho}(x) = n_0 \left(1+\sqrt{\frac{\alpha n_0}{6 D}} \,x\right)^{-2},
\end{equation}
it falls off much slower than the exponential profile (\ref{MFsteadysol}). For the continuous reaction-diffusion equation to be valid, it is necessary that the characteristic length scale $\sim (D/\alpha n_0)^{1/2}$ be much larger than the lattice constant $a$:
\begin{equation}\label{ineq1}
    \sqrt{\frac{D}{\alpha n_0}}\gg a.
\end{equation}
The average steady-state number of particles is
\begin{equation}\label{Nbarsteady1}
\bar{N} =\int_0^{\infty} dx\,\bar{\rho}(x)=\left(\frac{6 D n_0}{\alpha}\right)^{1/2},
\end{equation}
and we assume $\bar{N}\gg 1$. The MFT equations for this system can be safely derived from the exact master equation
for the multi-variate probability distribution by assuming that the typical number of particles on each lattice site is much larger than unity,
leading to the strong inequality $n_0 a \gg 1$  \cite{EK,MS2011}. We believe, however, that it is actually sufficient to require a weaker condition
$\bar{N}\gg 1$, alongside with the condition~\eqref{ineq1}. The MFT equations are \cite{MS2011}
\begin{eqnarray}
  \partial_t q &=& - \alpha q^2 e^{-2p} + D \partial_x\left(\partial_x q-2 q \partial_x p\right), \label{d11} \\
  \partial_t p &=& -\alpha q (e^{-2p}-1)- D \partial_x^2 p-D (\partial_x p)^2, \label{d21}
\end{eqnarray}
whereas
\begin{equation}
\label{Haman}
\mathcal{H}(q,p) = \frac{1}{2}\alpha q^2\left(e^{-2p}-1\right) -D \partial_x q \partial_x p
+D q\!\left(\partial_x p\right)^2
\end{equation}
is the Hamiltonian density  \cite{EK,MS2011}. The first term comes from the on-site annihilations, the second and third terms come from the diffusion. Correspondingly,
\begin{eqnarray}
\label{actionannih}
  &-&\ln {\mathcal P}(N) \simeq S = \int_{-\infty}^0  dt\, \int_0^{\infty} dx\,\left[D q (\partial_x p)^2+\frac{\alpha}{2} q^2 \left(1-e^{-2p}-2pe^{-2p}\right)\right].
\end{eqnarray}
As one can check, by performing rescalings described by Eq.~(\ref{transformation}) below and additional rescaling $q/n_0 \to q$,
\begin{equation}\label{scalingannih}
-\ln {\mathcal P}(N) = \bar{N} \,f\left(\frac{N}{\bar{N}}\right).
\end{equation}
Here $n_0$ only enters through $\bar{N}$, as for the random walkers who die individually, cf. Eq.~(\ref{scalingRW}).

As in Sec. \ref{var1}, we can calculate analytically the variance of the total number of ARWs in the steady state. We make the ansatz (\ref{q_exp}) and (\ref{p_exp}) in Eqs.~(\ref{d11}) and (\ref{d21})
and obtain, in the first order in $\lambda \ll 1$,
\begin{subequations}
\begin{align}
\label{q11}
&\left[\partial_t +2 \alpha \bar{\rho}(x) -D\partial^2_{x}\right] \,q_1 =  2 \alpha  \bar{\rho}^2(x) p_1-2 D \partial_x[\bar{\rho}(x)\partial_x p_1],\\
\label{p11}
&\left[\partial_t -2 \alpha \bar{\rho}(x) +D \partial^2_{x}\right] \,p_1=0.
\end{align}
\end{subequations}
As for the mortal SSEP, Eq.~(\ref{p11}) for $p_1$ is decoupled from that for $q_1$, and its solution suffices for computing the variance we are after.
Let us reverse and rescale time and transform the coordinate:
\begin{equation}\label{transformation}
\tau= -\frac{\alpha n_0 t}{6} \;\;\;\;\;\;\text{and}\;\;\;\; y= 1+\sqrt{\frac{\alpha n_0}{6 D}} \,x,
\end{equation}
so that Eq.~(\ref{p11}) becomes
\begin{equation}\label{pde2}
\partial_{\tau} p_1 +\frac{12 p_1}{y^2} = \partial^2_{y} p_1.
\end{equation}
We need to solve it for $1<y<\infty$ and $0<\tau<\infty$ subject to the boundary condition $p_1(y=1,\tau)=0$ and initial condition \begin{equation}\label{incond10}
p_1(y,\tau=0)=\theta(y-1) .
\end{equation}
The rescaled problem for $p_1(y,\tau)$ is parameter-free, and we solve it in Appendix \ref{A}.
Once $p_1(y,\tau)$ is found, we can calculate the
variance of $N$:
\begin{eqnarray}
\label{variancean}
 V_{\text{ARW}} &=& \int_{-\infty}^0  dt\, \int_0^{\infty} dx\,\left[2 D \bar{\rho} (\partial_x p_1)^2
 +2 \alpha \bar{\rho}^2 p_1^2\right] \nonumber \\
&=&2 \bar{N} \int_0^{\infty}  d\tau\, \int_1^{\infty} \frac{dy}{y^2} \left[(\partial_{y} p_1)^2+\frac{6 p_1^2}{y^2}\right].
\end{eqnarray}
As expected from Eq.~(\ref{scalingannih}), $V_{\text{ARW}}$ is proportional to $\bar{N}$. Using Eq.~(\ref{p1an}) of Appendix \ref{A} for $p(y,\tau)$, we can represent the
proportionality coefficient (a dimensionless number of order unity) as a quadruple integral. The integration over $\tau$ is elementary. The integration over $y$ is very tedious, but can be performed
explicitly with ``Mathematica". We evaluated the remaining double integral numerically, leading to
$ V_{\text{ARW}}\simeq 0.78 \bar{N}$.

As a check, we also solved Eq.~(\ref{pde2}) numerically in the region $1<y<L$ and $0<\tau<T$ with the boundary conditions $p_1(y=1,t)=0$ and $\partial_yp_1(y=L,t)=0$ and initial condition (\ref{incond10}), taking $L$ and $T$ sufficiently large. Then we used the numerical solution to compute the double integral in the second line of Eq.~(\ref{variancean}). The result comes quite close,  $V_{\text{ARW}}\simeq 0.77 \bar{N}$. As $V_{\text{ARW}} <\bar{N}$, the distribution ${\mathcal P}(N)$ is narrower than a Poisson distribution with the same mean.

\section{Summary and Discussion}

This work addressed the statistics of the total number of particles $N$ that are present at any chosen time in the steady state of a driven lattice gas composed of mortal diffusing particles. The formalism we used is that of the dissipative Macroscopic Fluctuation Theory (MFT).  For non-interacting random walkers who die individually, the MFT formulation of the
problem is exactly soluble and yields the expected Poissonian statistics of $N$ with the mean predicted by the simple reaction-diffusion equation (\ref{MFeq}). It also provides a fascinating and instructive visualization of large deviations of $N$ in the form of the optimal density history of the driven system conditional on $N$.  For interacting diffusing particles we calculated
the variance of the distribution of $N$, and found that the distribution is narrower than a Poisson distribution with the same mean.

The variance calculations that we showed here is a first step towards studying the complete statistics of $N$. Extending our perturbation theory for the MFT to higher orders in $\lambda$, one should be able to compute several higher moments of ${\mathcal P}(N)$, as
it has been recently done in the problem of melting of an Ising quadrant \cite{quadrant}. We also note that it is possible to compute the distribution of $N$ numerically by solving the full MFT equations with the proper boundary with 
the Chernykh-Stepanov iteration algorithm \cite{Stepanov}. This algorithm was originally developed for evaluating the probability distribution of large negative velocity gradients in the Burgers turbulence.  Later on it was used in studies of different types of large deviations in diffusive lattice gases, with and without on-site reactions \cite{EK,MS2011,KM_var,void,MVS,VMS,MVK}. This algorithm is much more computationally efficient than
microscopic stochastic simulations.

It would be very interesting, and challenging, to directly probe the tails of ${\mathcal P}(N)$, that are beyond the reach of the small-$\lambda$ perturbation theory. For the SSEP involving immortal particles, $\mu=0$, the limit of very large transferred  mass, in an infinite system, can be described by neglecting the term $-D\partial_x q\,\partial_x p$ in the Hamiltonian density \eqref{Ham}. The ensuing reduced MFT equations turn out to be exactly soluble \cite{MS2014,VMS}. Whether a similar reduction is possible in the problem of extreme statistics of the total number of interacting mortal particles is an open question.

On a more general note, understanding non-equilibrium systems requires, among other things, intuition which one acquires by learning from examples. The prototypical dissipative systems, considered in this work, are helpful in gaining such an intuition.

\section*{Acknowledgments}
This research was supported by grant No.\ 2012145 from the
United States--Israel Binational Science Foundation (BSF).

\appendix

\section{Calculating the variance for the SSEP}
\label{SSEP}

Here we evaluate the integral in Eq.~(\ref{variance}) for the SSEP. We start with calculating the integral
\begin{eqnarray}
\label{I1}
  I_1 &=& \int_{-\infty}^0  dt\, \int_0^{\infty} dx\,\sigma(\bar{\rho}) (\partial_x p_1)^2 \nonumber \\
  &=&  \int_0^{\infty} dx \, \int_{-\infty}^0  dt\, 2D n_0 \,e^{-\sqrt{\frac{\mu}{D}} x} \left(1-n_0\,e^{-\sqrt{\frac{\mu}{D}} x}\right) \left(-\frac{e^{-\frac{x^2}{2Dt}+2\mu t}}{\pi D t}\right) \nonumber\\
  &=& \frac{4n_0}{\pi} \int_0^{\infty} dx \,e^{-\sqrt{\frac{\mu}{D}} x} \left(1-n_0\,e^{-\sqrt{\frac{\mu}{D}} x}\right) \text{K}_0\left( \sqrt{\frac{4\mu}{D}}x\right),
\end{eqnarray}
where $\text{K}_0(\dots)$ is the modified Bessel function of the second kind. Evaluating the remaining integral, we obtain
\begin{equation}\label{I1result}
I_1=\sqrt{\frac{D}{\mu}} \,n_0 \left(\frac{4\sqrt{3}}{9}-\frac{2n_0}{\pi}\right).
\end{equation}
Now we evaluate the integral
\begin{eqnarray}
\label{I2}
  I_2 &=& \int_{-\infty}^0  dt\, \int_0^{\infty} dx\,\mu \bar{\rho}p_1^2 \nonumber \\
  &=& \int_{-\infty}^0  dt\, \int_0^{\infty} dx \, \mu n_0 \,e^{-\sqrt{\frac{\mu}{D}} x+2 \mu t} \,\text{erf}^2\,\left(\frac{x}{\sqrt{-4Dt}}\right)\nonumber\\
  &=& \sqrt{\frac{4D}{\mu}} \,n_0 \int_0^{\infty} dz \sqrt{z}\, e^{-2z} \int_0^{\infty} du\,e^{-2\sqrt{z} \,u} \,\text{erf}^2\,u.
\end{eqnarray}
With some patience, this double integral can be evaluated using the book of integrals \cite{Prudnikov}, and the result is \cite{noneed}
\begin{equation}\label{I2result}
I_2= \sqrt{\frac{D}{\mu}} \,n_0 \left(1-\frac{4\sqrt{3}}{9}\right).
\end{equation}
Summing up $I_1$ and $I_2$ we obtain Eq.~(\ref{varianceSSEP}).

\section{Finding $p_1(y,\tau)$ for annihilating random walkers}
\label{A}
To find the spectrum of the problem and the proper eigenfunctions, we make the ansatz
$p_1(y,\tau)=\psi(y,\Gamma) e^{-\Gamma^2 \tau}$ and arrive at the equation
\begin{equation}\label{Sch}
\frac{d^2 \psi(y,\Gamma)}{dy^2} +\left(\Gamma^2-\frac{12}{y^2}\right) \psi(y,\Gamma)=0
\end{equation}
that we need to solve with  the boundary condition
$\psi(y=1,\Gamma)=0$. Equation~(\ref{Sch}) is the Shr\"{o}dinger equation for a quantum particle with
energy $\Gamma^2$ in the potential
\begin{equation}
\label{potential}
V(y) =
\begin{cases}
\frac{12}{y^2},   & y>1,\\
\infty,  & y\leq 1.
\end{cases}
\end{equation}
The spectrum of the problem is continuous, $0<\Gamma<\infty$. Two linearly independent solutions of Eq.~(\ref{Sch}) can be chosen  as
\begin{equation}\label{psi1}
\psi_1(y,\Gamma) =\frac{\Gamma  y \left(\Gamma ^2
   y^2-15\right) \sin (\Gamma
   y)+3 \left(2 \Gamma ^2
   y^2-5\right) \cos (\Gamma
   y)}{\Gamma ^3 y^3}
\end{equation}
and
\begin{equation}\label{psi2}
\psi_2 (y,\Gamma) =\frac{\Gamma  y \left(\Gamma ^2
   y^2-15\right) \cos (\Gamma
   y)-3 \left(2 \Gamma ^2
   y^2-5\right) \sin (\Gamma
   y)}{\Gamma ^3 y^3}.
\end{equation}
At $\Gamma\to \infty$ (or at $y\to \infty$) these solutions become $\sin (\Gamma y)$ and
$\cos (\Gamma y)$ as expected.  All the eigenfunctions, vanishing at $y=1$, can be written as
\begin{equation}\label{eigenfunc}
\psi(y,\Gamma)=a(\Gamma) \phi(y,\Gamma),
\end{equation}
where
\begin{equation}\label{lincombbasic}
\phi(y,\Gamma)= \psi_1 (y,\Gamma) \psi_2(1,\Gamma)- \psi_1 (1,\Gamma) \psi_2 (y,\Gamma)
\end{equation}
and $a(\Gamma)$ is a yet undetermined amplitude. The eigenfunctions (\ref{eigenfunc}) are orthogonal, and they can be normalized as follows:
\begin{equation}\label{normalization}
\int_1^{\infty} dy\, \psi (y,\Gamma) \,\psi (y, \Gamma^{\prime})=\delta (\Gamma-\Gamma^{\prime}),
\end{equation}
where $\delta$ is Dirac's delta function. As a result,
\begin{equation}\label{a}
a^{-2}(\Gamma)=\int_0^{\infty} d\Gamma^{\prime} \int_1^{\infty} dy \,\phi(y,\Gamma)\,\phi(y,\Gamma^{\prime}).
\end{equation}
Evaluating this double integral with a help of ``Mathematica", we obtain
\begin{equation}\label{afinal}
a(\Gamma)=\sqrt{\frac{2}{\pi}}\,\frac{\Gamma^3}{\sqrt{\Gamma^6+6\Gamma^4+45 \Gamma^2+225}}.
\end{equation}
The solution for $p_1(y,\tau)$ can be written as
\begin{equation}\label{p1an}
p_1(y,\tau)=\int_0^{\infty} d \Gamma A(\Gamma) \,\psi (y,\Gamma) \,e^{-\Gamma^2 \tau},
\end{equation}
where $A(\Gamma)$ is the projection of the initial condition (\ref{incond10}) on the normalized eigenfunctions (\ref{eigenfunc}). That is, $A(\Gamma)=\int_1^{\infty} dy\, \psi(y,\Gamma)$.
This integral can be also evaluated with ``Mathematica", resulting in a tedious formula
\begin{footnotesize}
\begin{eqnarray}
\nonumber
  A(\Gamma) &=& \frac{4 \Gamma^3+90 \Gamma +6 \text{Ci}(\Gamma) \left[3 \left(5-2 \Gamma ^2\right) \sin \Gamma
   +\Gamma  \left(\Gamma ^2-15\right) \cos \Gamma \right]-3  \Gamma \left(\Gamma ^2-15\right)
   [\pi -2 \text{Si}(\Gamma )]  \sin \Gamma -9 \left(2 \Gamma ^2-5\right) [\pi -2
   \text{Si}(\Gamma )]  \cos \Gamma }{2 \sqrt{2 \pi }
   \, \Gamma \, \sqrt{\Gamma ^6+6 \Gamma ^4+45 \Gamma ^2+225}},
\end{eqnarray}
\end{footnotesize}

\noindent where
$$
\text{Si}(z) =\int_0^z \frac{\sin y}{y}\,dy\;\;\;\text{and}\;\;\;\text{Ci}(z) =-\int_z^{\infty} \frac{\cos y}{y}\,dy
$$
are the sine and cosine integrals, respectively. Now $p_1(y,\tau)$ in Eq.~(\ref{p1an}) is fully determined in terms of a single integral over $\Gamma$.


\begin{thebibliography} {99}

\bibitem{Spohn}

H. Spohn, {\it Large Scale Dynamics of Interacting Particles} (Springer-Verlag, New York, 1991).

\bibitem{KL99}

C. Kipnis and C. Landim, {\it Scaling Limits of Interacting Particle Systems} (Springer-Verlag, New York,  1999).

\bibitem{SZ95}

B. Schmittmann and R. K. P. Zia, \textit{Statistical Mechanics of Driven Diffusive Systems},
in: {\it Phase Transitions and Critical Phenomena}, Vol.\ 17, eds.\ C. Domb and J. L. Lebowitz
(Academic Press, London, 1995).

\bibitem{S00}

G. Sch\"utz, \textit{Exactly Solvable Models for Many-Body Systems
  Far From Equilibrium}, in {\it Phase Transitions and Critical Phenomena},
  Vol.\ 19, eds.\ C. Domb and J. L. Lebowitz (Academic Press, London, 2000).

\bibitem{D07}

B. Derrida, J. Stat. Mech. P07023 (2007).


\bibitem{BE07} R. A. Blythe and M. R. Evans, J. Phys.\ A {\bf 40}, R333
  (2007).

\bibitem{JL2014} G. Jona-Lasinio, J. Stat. Mech. P02004 (2014).


\bibitem{JonaLasinio93}

G. Jona-Lasinio, C. Landim, and M.E. Vares, 
Probability Theory and Related Fields \textbf{97}, 339 (1993).

\bibitem{Basile}

G. Basile and G. Jona-Lasinio, 
Int. J. Mod. Phys. B \textbf{18} , 479 (2004).

\bibitem{BodineauLag}

T. Bodineau and M. Lagouge, 
J. Stat. Phys.  \textbf{139}, 201 (2010).

\bibitem{Hurtado}

P.I. Hurtado, A. Lasanta, and A. Prados, Phys. Rev. E \textbf{88}, 022110 (2013).

\bibitem{Frisch}

U. Frisch, {\it Turbulence: The Legacy of A. N. Kolmogorov} (Cambridge University Press, Cambridge, UK, 2001).

\bibitem{Poeschel1}

{\it Granular Gases}, eds.\ T. P\"{o}schel and S. Luding
(Springer-Verlag, Berlin, 2001).

\bibitem{Poeschel2}

{\it Granular Gas Dynamics}, eds.\ T. P\"{o}schel and N. Brilliantov
(Springer-Verlag, Berlin, 2003).

\bibitem{MFTreview}

L. Bertini, A. De Sole, D. Gabrielli, G. Jona-Lasinio, and C. Landim, arXiv:1404.6466.

\bibitem{Lebowitz}

A. De Masi, P. Ferrari, and J. Lebowitz, 
Phys. Rev. Lett. \textbf{55}, 1947 (1985);  
J. Stat. Phys. \textbf{44}, 589 (1986).


\bibitem{Mikhailov}

A. S. Mikhailov, \textit{Foundations of Synergetics I. Distributed Active
Systems} (Springer-Verlag, Berlin, 1990).

\bibitem{EK}

V. Elgart and A. Kamenev, Phys. Rev. E 70, 041106 (2004).


\bibitem{MS2011}

B. Meerson and P.V. Sasorov, Phys. Rev. E \textbf{83}, 011129 (2011).

\bibitem{Kurchan}

J. Tailleur, J. Kurchan, and V. Lecomte, Phys. Rev. Lett. \textbf{99}, 150602 (2007).


\bibitem{DG2009b}

B. Derrida and A. Gerschenfeld, J. Stat. Phys. \textbf{137}, 978 (2009).

\bibitem{MR} B. Meerson and S. Redner, J. Stat. Mech. (2014) P08008.

\bibitem{KM_var}

P. L. Krapivsky and B. Meerson,  Phys. Rev. E \textbf{86}, 031106 (2012).

\bibitem{scalesigma} While performing the rescaling, we assumed that $\sigma(q)=D \,\tilde{\sigma} (q)$,
where $\tilde{\sigma}(q)$ has the dimensions of $q$.

\bibitem{quadrant}

P. L. Krapivsky, K. Mallick,  and T. Sadhu, J. Phys. A: Math. Theor. \textbf{48}, 015005 (2015).

\bibitem{Stepanov}

A. I. Chernykh and M. G. Stepanov, Phys. Rev. E \textbf{64}, 026306 (2001).

\bibitem{void}

P.L. Krapivsky, B. Meerson, and P.V. Sasorov, J. Stat. Mech. (2012) P12014.

\bibitem{MVS}

B. Meerson, A. Vilenkin, and P.V. Sasorov,  Phys. Rev. E \textbf{87}, 012117 (2013).

\bibitem{VMS}

A. Vilenkin, B. Meerson, and P.V. Sasorov,  J. Stat. Mech. (2014) P06007.

\bibitem{MVK}

B. Meerson, A. Vilenkin, and P.L. Krapivsky, Phys. Rev. E \textbf{90}, 022120 (2014).

\bibitem{MS2014}

B. Meerson and P.V. Sasorov, Phys. Rev. E  \textbf{89}, 010101(R) (2014).

\bibitem{Prudnikov}

A. P. Prudnikov, I.A. Brychkov, and O. I. Marichev, {\it Integrals and Series: Special Functions}
(CRC Press, London, 1998).

\bibitem{noneed}

In fact, the calculation of $I_2$ is unnecessary. This is because, when the term $2n_0/\pi$ in the parentheses of Eq.~(\ref{I1result}) is disregarded, the resulting variance must coincide with that for the non-interacting random walkers. Hence, without
the $2n_0/\pi$ term, $I_1+I_2$ is given by Eq.~(\ref{varRW}). What is left then is to calculate the contribution
of the  $2n_0/\pi$ term.

	



\end{thebibliography}
\end{document}